\documentclass[twocolumn,showpacs,preprintnumbers,amsmath,amssymb]{revtex4}

\usepackage{graphicx}
\usepackage{dcolumn}
\usepackage{bm}
\newcommand{\ket}[1]{\big|#1\big>}
\newcommand{\bra}[1]{\big<#1\big|}

\newcommand{\proj}[1]{\ket{#1}\bra{#1}}

\begin{document}

\title{Bound entanglement maximally violating Bell inequalities: \\
quantum entanglement is not equivalent to quantum security
}

\author{Remigiusz Augusiak}
 \email{remik@mif.pg.gda.pl}
\author{Pawel Horodecki}%
 \email{pawel@mif.pg.gda.pl}
\affiliation{%
Faculty of Applied Physics and Mathematics \\
Gda\'nsk University of Technology, Gda\'nsk, Poland
}%

\date{\today}

\begin{abstract}
It is shown that Smolin four-qubit bound entangled states
[Phys. Rev. A, {\bf 63} 032306 (2001)] can
maximally violate two-setting Bell inequality similar to standard
CHSH inequality. Surprisingly this entanglement does not allow for
secure key distillation, so neither entanglement nor violation of
Bell inequalities implies quantum security. It is also pointed out how
that kind of bound entanglement can be useful in reducing
communication complexity.
\end{abstract}

\maketitle
Quantum entanglement is one of the most intriguing phenomenon
within quantum physics. The pure state of composite quantum system
is called to be entangled if it is impossible to describe its
subsystems by pure states. Historically  importance of quantum
entanglement has been recognized  by Einstein, Podolsky and Rosen
(EPR) \cite{EPR} and Schr\"odinger  \cite{Sch}. The so-called EPR
paradox has started long debate whether local realistic theories
can simulate quantum mechanics. The  well-known Bell theorem
\cite{Bell} says that such a simulation is, in general,
impossible. On the other hand, entanglement has been found as an
important resource in quantum information theory \cite{Ksiazki}
being an essential ingredient of coding, entanglement based
cryptographic scheme \cite{Artur},  quantum dense coding
\cite{geste}, quantum teleportation \cite{teleportacja} and
quantum computing \cite{Shor}.

To overcome the problem of noisy entanglement \cite{Werner} the
idea of entanglement distillation has been invented \cite{dist1}
which is useful in quantum privacy amplification \cite{QPA}. While
any entangled two-qubit (or qubit-qutrit) state can be distilled
to a  singlet form \cite{dist} this is not true in general and
this fact reflects the existence of  so called bound entanglement
phenomenon \cite{bound}.  This is very weak type of entanglement
which can not serve in dense coding or teleportation. However, in
multipartite case (see \cite{UPB}) can be useful for remote
quantum information concentration \cite{conc} (bipartite)
activation \cite{activ} and (multipartite) superactivation
\cite{superactivAB}, classical cryptographic key distillation
\cite{boundkey} or nonadditivity of  quantum channels with
multiple receivers \cite{multiple}. Remarkably bipartite BE states
have not been reported to violate any Bell inequalities so far
(see \cite{WW,Zukowski,Acin}). For multipartite case the seminal
result has been obtained by D\"ur \cite{Dur} who showed  that some
multiqubit BE states violate two-settings Bell inequalities. The
quite interesting question has arose: {\it what is the minimal
size of quantum system (in terms of subsystems) that admits bound
entanglement to violate Bell inequalities ?}.

It is important question since violation of Bell inequalities by
entangled states seems to imply  that the entanglement is useful
for some classical tasks. In particular, it has been shown that
violation of wide class of Bell inequalities is equivalent to
possibility of reduction of communication complexity by the
corresponding states \cite{complexity1,complexity2}. There is even
a conjecture following cryptography analysis (see \cite{crypto})
that violation of Bell inequalities is an indicator of usefulness
of entanglement in general \cite{GisinUsefulness}.

The minimal number of bound entangled qubits violating the
inequalities in D\"ur scheme was $N=8$. This limit has been pushed
down by Kaszlikowski {\it et al.} to \cite{Kaszlikowski} $N\geq 7$
with help of three apparatus settings per site and further by Sen
{\it et al.} \cite{Sen} to $N\geq 6$ with help of so--called
functional Bell inequalities \cite{functional}. The relation of
the results to bipartite distillability has been analyzed in
details in \cite{Acin}.

In the present paper we show that there are BE states that violate
{\it maximally} standard Bell inequalities, i.e., with two setting
per site \cite{WW,Zukowski} for $N=4$. The inequality is very
similar to standard CHSH inequality for two qubits \cite{CHSH}.
Moreover, the robustness of the considered bound entanglement is
comparable to that of entangled two-qubit Werner states. Note that
maximal violation of Bell inequalities by mixed states has already
been known \cite{BMR}. However, this is the first time when such a
violation is reported for bound entangled states.

Surprisingly, despite maximal violation, {\it of the same kind as
for pure GHZ state,} the considered states do not allow for secure
key distillation as it was for bipartite BE states from Ref.
\cite{boundkey}. This is a striking feature: knowing about Ekert
protocol for bipartite states \cite{Artur} one might expect that
violation of Bell inequalities is always indicator of secure key distillation.
A fundamental implication of our result is the
conclusion that being a precondition for quantum cryptography \cite{Curty} entanglement is {\it not} sufficient for the latter.

The BE states we shall show to violate Bell inequalities are
Smolin states \cite{Smolin}. They have a special property: if the
four particles are far apart no entanglement between any
subsystems can be distilled. If however two particles are in the
some location - it is possible to create maximal entanglement
between remaining two particles by means of LOCC operations.
Detailed analysis has shown further interesting aspects: the
entanglement cost of such states corresponds just to a singlet
state \cite{Tsu-Chieh}.

Using the result we also show, {\it via} results
of Ref. \cite{complexity1,complexity2} that considered bound entanglement
can serve to reduce communication complexity.

{\it Noisy Smolin states .-} Lets consider the following
four-qubit mixed state ($\varrho$ defined on space
$(\mathbb{C}^{2})^{\otimes 4}$):
\begin{equation}\label{2.1}
\varrho_{ABCD}^{S}(p)=\varrho^{S}(p)\equiv
(1-p)\frac{I^{\otimes4}}{16} +p\rho^{S},
\end{equation}
where $I$ stands for identity on one-qubit space, while $\rho^{S}$
is the four-qubit bound entangled state introduced by Smolin
\cite{Smolin} and is defined through the relationship
%
\begin{eqnarray}
\rho^{S}_{ABCD}=\rho^{S}=\frac{1}{4}\sum_{i=1}^{4}\proj{\psi_{i}}\otimes\proj{\psi_{i}},
\label{2.2}
\end{eqnarray}
%
where two-particle states
%
$\ket{\psi_{1(2)}}=\frac{1}{\sqrt{2}} (\ket{00}\pm\ket{11})$,
%
$\ket{\psi_{3(4)}}=\frac{1}{\sqrt{2}} (\ket{01}\pm\ket{10})$
%
form the so-called Bell basis. It is worth noticing that the
states (\ref{2.1}) are fully permutationally invariant, since they
can be written in the form
\begin{equation}\label{}
\varrho^{S}(p)=\frac{1}{16}\biglb(I\otimes I\otimes I\otimes I+
p\sum_{i=1}^{3}\sigma_{i}\otimes\sigma_{i}\otimes\sigma_{i}\otimes\sigma_{i}\bigrb)
\end{equation}
with $\sigma_{i}$ denoting the standard Pauli matrices.
 One can see that for $p=1/3$ the state
$\varrho^{S}(p)$ is separable. Indeed, for such value of $p$ we
can rewrite (\ref{2.1}) as
\begin{equation}\label{2.5}
\varrho^{S}({\textstyle\frac{1}{3}})=\frac{1}{6}\sum_{i=1}^{3}\sum_{s=-}^{+}\varrho_{i}^{(s)}\otimes\varrho_{i}^{(s)},
\end{equation}
where $\varrho_{k}^{(\pm)}$ stand for two-qubit separable mixed
states (see \cite{Horodeccy1}):
\begin{equation}\label{2.6}
\varrho_{k}^{(\pm)}=\frac{1}{2}\biglb(P_{k}^{(+)}\otimes
P_{k}^{(\pm)}+P_{k}^{(-)}\otimes P_{k}^{(\mp)}\bigrb), \quad
k=1,2,3,
\end{equation}
$P_{k}^{(\pm)}$ denotes projectors corresponding to the
eigenvectors of $\sigma_{k}$ with eigenvalues $\pm1$. Since by
LOCC we can add some noise to the state, the above fact implies
that for all $p\le 1/3$ the state $\varrho$ is separable. On the
other hand for $p> 1/3$ the state is bound entangled. Indeed, it
is easy to see that for this region the state
 violates PPT separability criterion
\cite{Peres} if we transpose indices corresponding to single qubit
ie. against any of the cuts: $A|BCD$, $B|ACD$ etc. Thus the state
is entangled. It is  bound entangled since the state maintains the
property of original $\rho^{S}$: it is separable against bipartite
symmetric cuts like $AB|CD$, $BC|AD$, etc., which makes sure that
no maximal entanglement between any subsystems can be distilled.

Below we shall prove that for $p>1/\sqrt{2}$ the state violates
Bell inequalities introduced in Refs. \cite{WW,Zukowski}.

{\it Violation of Bell inequalities .-}

In the corresponding scenario each of the $N$ parties
corresponding to index $j$  ($j=1,2,...,N$) can choose between two
dichotomic observables $(O_{j}^{{k}_{j}})$, $k_{j}=1,2$. The set
of $2^{2^{N}}$ Bell inequalities is \cite{WW,Zukowski}:
\begin{equation}\label{3.1}
|\sum_{s_{1},..,s_{N}}^{\pm1}
S(s_{1},..,s_{N})\sum_{k_{1},..,k_{N}}^{1,2}s_{1}^{k_{1}-1}..
s_{N}^{k_{N}-1}E(k_{1},..,k_{N})|\le 2^{N},
\end{equation}
where correlation function
$E$ is defined through relation (average over many
runs of experiment)
\begin{equation}\label{3.2}
E(k_{1},\ldots,k_{N})=\big{\langle}\prod_{j=1}^{N}
O^{(j)}_{{k_{j}}} \big{\rangle}_{avg}.
\end{equation}
Trying to predict  the above (experimental)
average local hidden variables (LHV) theories offer
its calculation as an integral over probabilistic measure
on space of  ,,hidden parameters''. The measure correspond
to classical states. In quantum mechanical  regime
the observables depend on vector parameters
\begin{equation}
O^{(j)}_{k_{j}}=\hat{n}^{(j)}_{k_{j}}\roarrow{\sigma}
\end{equation}
and the corresponding average for given
quantum state $\varrho$ is calculated as follows:
\begin{equation}\label{3.3}
E_{QM}(k_{1},\ldots,k_{N})(\varrho)=\textrm{Tr}\left[\varrho
O_{k_{1}}^{(1)} \otimes \ldots \otimes O_{k_{N}}^{(N)}\right].
\end{equation}
One can see that for the following non-trivial sign function:
\begin{eqnarray}\label{signB}
&&S(+,+,-,-)=S(+,-,+,-)\nonumber\\
&&=S(-,+,+,-)=S(-,-,-,-)=-1,
\end{eqnarray}
where $\pm$ stands for $\pm 1$ (for other cases sign function
equals to unity), and for $N=4$ we can derive the following Bell-inequality:
\begin{equation}\label{nBella}
|E(1,1,1,1)+E(1,1,1,2)+E(2,2,2,1)-E(2,2,2,2)| \le 2.
\end{equation}
This inequality can be also derived very easily using the same
technique as in standard CHSH inequality \cite{CHSH}. We keep the
above derivation for purposes of further analysis.

Subsequently, for the state given by (\ref{2.2}) we choose the
following observables:
\begin{eqnarray}\label{3.5}
&&\hat{n}^{(i)}_{1}=\hat{x},\quad
\hat{n}^{(i)}_{2}=\hat{y}, \quad i=1,2,3 \nonumber\\
&&\hat{n}^{(4)}_{1}=\frac{\hat{x}+\hat{y}}{\sqrt{2}},\quad
\hat{n}^{(4)}_{2}=\frac{\hat{x}-\hat{y}}{\sqrt{2}}.
\end{eqnarray}
In virtue of the above equations, one gets the violation of the
Bell inequality \eqref{nBella}:
\begin{eqnarray}
&&E_{QM}(1,1,1,1)(\varrho_{S}(p))+E_{QM}(1,1,1,2)(\varrho_{S}(p))\nonumber \\
&&+E_{QM}(2,2,2,1)(\varrho_{S}(p))-E_{QM}(2,2,2,2)(\varrho_{S}(p)) \nonumber \\
&&=2\sqrt{2}p, \label{2s2p}
\end{eqnarray}
which gives violation for any $p\in( \frac{1}{\sqrt{2}},1]$. Below
we shall show that for $p=1$, i.e., for original Smolin states
$\varrho^{S}(p)$ \cite{Smolin}  the above violation is maximal,
i.e., there is no other quantum state that can make RHS of the
above equation greater than $2 \sqrt{2}$. To this aim we use the
method \cite{Tsirelson} applied for CHSH inequality, namely, we
shall calculate the second power of  the Bell operator:
\begin{eqnarray}
&&\hspace{-0.75cm}B=O^{(1)}_{1}\otimes O^{(2)}_{1} \otimes O^{(3)}_{1}\otimes O^{(4)}_{1} \nonumber \\
&&+O^{(1)}_{1}\otimes O^{(2)}_{1} \otimes O^{(3)}_{1}\otimes O^{(4)}_{2} \nonumber \\
&&+O^{(1)}_{2}\otimes O^{(2)}_{2} \otimes O^{(3)}_{2}\otimes O^{(4)}_{1} \nonumber \\
&&-O^{(1)}_{2}\otimes O^{(2)}_{2} \otimes O^{(3)}_{2}\otimes
O^{(4)}_{2}.
\end{eqnarray}
Since the observables are dichotomic (i.e., one has
$(O^{(j)}_{k_{j}})^{2}=I$) we get that
\begin{eqnarray}
&& \hspace{-0.3cm}B^{2}=4I\otimes I \otimes I \otimes I +
(O^{(1)}_{1}O^{(1)}_{2}\otimes O^{(2)}_{1}O^{(2)}_{2}
\otimes O^{(3)}_{1}O^{(3)}_{2}  \nonumber \\
&&-O^{(1)}_{2}O^{(1)}_{1}\otimes O^{(2)}_{2}O^{(2)}_{1} \otimes
O^{(3)}_{2}O^{(3)}_{1})\otimes[O^{(4)}_{1},O^{(4)}_{2}],
\end{eqnarray}
where the last term stands for commutator. Again, by virtue of
dichotomic character  of the observables involved, this can be
written as $B^{2}=4I+(X-X')\otimes(Y-Y')$ with all $X,X',Y,Y'$
having operator norm not greater than one. This gives the estimate
$||B^{2}||\leq 4+(||X||+||X'||)(||Y|| +||Y'||)\leq 8$ and thus the
spectrum of $B$ lies within the interval $[-2\sqrt{2},2\sqrt{2}]$.
This immediately implies that RHS of (\ref{2s2p}) achieves its
maximum in $2\sqrt{2}$.

Remarkable that both  separability and violation of Bell
inequalities (\ref{nBella}) is in the same regime as in two-qubit
Werner states \cite{Werner} if we write them in the form
$\varrho^{W}(p) = (1-p) \frac{1}{4}  I \otimes I +
p|\Phi_{-}\rangle \langle\Phi_{-}|$ and consider CHSH inequality.
Indeed, $\varrho^{W}(p)$ is known to (i) be entangled for $p \in
(\frac{1}{3},1]$ \cite{Werner,huge} and (ii) to violate Bell-CHSH
inequality for  $p \in (\frac{1}{\sqrt{2}},1]$ \cite{BellMPRH}.

{\it Communication complexity .-}
In this section we analyze the Smolin state in context of
communication complexity problems. To aim this let us consider the
general class of problems, proposed by Brukner \textit{et al.}
\cite{complexity1}:
\begin{itemize}
    \item The $i$-th party receives a two-bit input string $(x_{i},y_{i})$
($x_{i}\in\{1,2\},\quad y_{i}\in\{-1,1\}$). Values of $y_{i}$ are
chosen randomly, whereas values of $x_{i}$ are chosen according to
probability distribution $Q(x_{1},\ldots,x_{N})$ for which one can
find real--valued function $g(x_{1},\ldots,x_{N})$ such that
\begin{equation}\label{4.1}
Q(x_{1},\ldots,x_{4})=\frac{g(x_{1},\ldots,x_{N})}
{\sum_{x_{1},\ldots,x_{N}=1}^{2}|g(x_{1},\ldots,x_{N})|},
\end{equation}
    \item Each party is
allowed to distribute one classical bit of information.
    \item The main
goal of each party is to find correct value of the function
\begin{equation}\label{4.2}
f=y_{1}\cdot\ldots\cdot y_{N}S[g(x_{1},\ldots,x_{N})],
\end{equation}
where $S[g]=g/|g|=\pm1$ is the sign function of $g$ and
$f\in\{-1,1\}$. The success is achieved when all parties get the
correct value of $f$. Their joint task is to maximize the
probability of success.
\end{itemize}

Following the broad class of quantum protocols
\begin{itemize}
    \item After receiving $x_{i}=1$ $(x_{i}=2)$, the $i$-th party
    set its apparatus to measurement of a dichotomic observable
    $O_{1}^{(i)}$ $(O_{2}^{(i)})$. Values obtained in such measurements
    are an elements of the set $\{-1,1\}$ and will be denoted by
    $a_{i}$. Each party distribute one bit of classical
    information $e_{i}=a_{i}\cdot y_{i}$.
    \item Finally, all parties compute the product \linebreak$y_{1}\cdot\ldots\cdot y_{N}\cdot
a_{1}\cdot\ldots\cdot a_{N}$ and put it as a value of function $f$
\end{itemize}
it is proven in \cite{complexity1} that probability of success in
quantum case is higher than in classical one iff for given
entangled state one of the following Bell inequality for
correlation function
\begin{equation}\label{4.3}
\sum_{x_{1},\ldots,x_{N}=1}^{2}g(x_{1},\ldots,x_{N})E(x_{1},\ldots,x_{N})\le
B(N)
\end{equation}
is violated. In particular, the above class contains Bell
inequalities given by (\ref{3.1}) with $B(N)=2^{N}$ and
\begin{equation}\label{4.7}
g(x_{1},\ldots,x_{N})=\sum_{s_{1},..,s_{N}}^{\pm1}S(s_{1},..,s_{N})s_{1}^{x_{1}-1}\cdot\ldots\cdot
s_{N}^{x_{N}-1}.
\end{equation}

It is worth noticing that the probability $P$ of success in case
of classical protocols is estimated as follows (for proof see
\cite{complexity1})
\begin{equation}\label{Prob}
P\le \frac{1}{2} \left( 1+\frac{B(N)}{\sum|g(x_{1},\ldots,x_{N})|}
\right).
\end{equation}
On the other hand the above inequalities are equivalent to Bell
inequalities (\ref{4.3}), i.e., are violated iff the inequalities
(\ref{4.3}) are violated. Therefore, one can see that violation of
Bell inequality implies violation of respective inequality
(\ref{Prob}) and can result in higher probability of success in
case of quantum entanglement.

To show that Smolin state is useful to reduce communication
complexity it is sufficient to consider the function $g$ given by

\begin{eqnarray}\label{functionG}
&&\hspace{-0.5cm}g(x_{1},..,x_{4})=4\sqrt{2}\,\cos[(x_{1}-x_{2})\textstyle\frac{\pi}{2}]
\sin\textstyle[(\frac{3}{2}(-1)^{x_{4}}-x_{3})\frac{\pi}{2}]\nonumber\\
&&\hspace{0.5cm}+4\sqrt{2}\,\cos\textstyle[(x_{1}+x_{2})\frac{\pi}{2}]
\sin\textstyle[(\frac{3}{2}(-1)^{x_{4}}+x_{3})\frac{\pi}{2}]
\end{eqnarray}
and to put $B(4)=16$. Note that function (\ref{functionG}) is
equivalent to the function obtained after substitution of
(\ref{signB}) to (\ref{4.7}) for $x_{i}\in\{1,2\}$.

By virtue of \eqref{Prob}, we infer that maximal probability
achievable in any classical protocol is $P_{max}^{C}=3/4$, whereas
in quantum protocol is
$P_{max}^{Q}=\frac{1}{2}(1+\frac{1}{\sqrt{2}})$.

{\it Discussion and conclusions .-} Bound entanglement is quite
unique type of  entanglement that does not posses property of
distillability. Nevertheless it can be useful to perform some
quantum tasks. It is located in a sense in between usual free
,,strong'', entanglement and  separability. As such it represents
the region of quantumness where natural limits of classical
theories can be tested. One of the fundamental questions are
limits of local hidden variables theories - so far it has been
known that they are excluded for BE states with number of qubits
not less that six. We show that violation of hidden variable model
test is possible by four--qubit system. In particular four--qubit
Smolin states can violate CHSH-like Bell inequalities {\it
maximally}. This is the first time when such violation is proved
for BE states. Moreover, the result pushes down the number of
qubits needed for BE to violate Bell inequalities: from $6$ to
$4$. If we consider the family of {\it two settings inequalities}
(which are most easy to implement) Bell inequalities the present
offers significant progress from $N=8$ \cite{Dur} to $N=4$.

Even more striking are implications of the present result as far
as quantum security is concerned: for the first time we have
example showing that {\it neither Bell inequalities violation nor
entanglement itself implies quantum security}. Indeed, the states
are separable under any symmetric bipartite cut, which means
(following analysis of Ref. \cite{Curty}) that no secure
correlations can be distilled between any two groups of two people
in the scheme. By full permutational symmetry of the state this
means that no secure key between four people can be distilled.
Moreover, from the result of the present paper it follows that
possibilities of performing two important quantum information
tasks are {\it not} equivalent. Indeed, possibility of generation
of quantum secure key is not equivalent to possibility of reducing
of communication complexity. The additional surprise is that at
the same time the states violate the considered Bell inequality in
the same way as (maximally entangled) pure GHZ state.

%

{\it Acknowledgements .-} RA thanks to Maciej Demianowicz for
stimulating discussions. PH thanks Ryszard, Michal and Karol Horodecki for
helpful disussions. The work is supported by European Union under
grant RESQ No. IST-2001-37559 and grant QUPRODIS No.
IST-2001-38877, and by Polish Ministry of Science under grant  No.
PB2-MIN-008/P03/2003.

\end{document}